\documentclass[aps,pra,twocolumn]{revtex4}
\usepackage{graphicx}    
\usepackage{amsmath}

\begin{document}

\title{Bose-Einstein Condensation and Supersolids}

\author{Moorad Alexanian$^{1}$ and Vanik E. Mkrtchian$^{2}$ }
\affiliation{$^{1}$Department of Physics and Physical Oceanography, University of North Carolina Wilmington, Wilmington, NC 28403-5606, USA\\$^{2}$Institute for Physical Research, Armenian Academy of Sciences, Ashtarak 0203, Republic of Armenia}

\date{\today}

\begin{abstract}

\noindent \textbf{Abstract:} We consider interacting Bose particles in an external potential. It is shown that a Bose-Einstein condensate is possible at finite temperatures that describes a supersolid in three dimensions (3D) for a wide range of potentials in the absence of an external potential. However, for 2D, a self-organized supersolid exists for finite temperatures provided the interaction between bosons is nonlocal and of infinitely long-range. It is interesting that in the absence of the latter type of potential and in the presence of a lattice potential, there is no Bose-Einstein condensate and so in such a case, a 2D supersolid is not possible at finite temperatures. We also propose the correct Bloch form of the condensate wave function valid for finite temperatures, which may be used as the correct trial wave function.
\end{abstract}

\maketitle {}

\section{Introduction}

The existence of a Bose-Einstein condensate (BEC) in an ideal three-dimensional (3D) quantum Bose gas served as a useful physical concept in the study of theoretical models for superfluidity, which are usually associated with the presence of a BEC. It is interesting that the physical realization of a BEC in a dilute gas has given impetus to the study of BEC for its own sake with a view of understanding many-body systems \cite{EJM96}. In addition, the confinement of photons and molecules in thermal equilibrium in an optical cavity reveals a BEC even for photons \cite{KS10}. The existence of superflow \cite{EK04} in solid helium $^4 \textup{He}$ has stimulated the search of a BEC in solid helium thus establishing the existence of BECs in all three states of matter--gas, liquid, and solid. The emergence of a self-organized supersolid phase, both a superfluid with crystalline order simultaneously, formed by a BEC coupled to an optical cavity has been observed \cite{BG10}.

The quantum phase transition describing the supersolid, which is associated with a spontaneous broken spatial symmetry, is in quantitative agreement with the Dicke model of superradiance and is driven by an infinitely long-range interaction between the condensed atoms \cite{BG10}. A nonlocal potential is found to favor a crystalline BEC for the ground state of two-dimensional interacting bosons \cite{LLL11}. Numerical techniques have been used to predict a novel supersolid phase for an ensemble of Rydberg atoms in the dipole-blocked regime confined to two dimensions, interacting via a repulsive dipole potential softened at short distances \cite{CJB10}.  It is claimed that the superfluid droplet-crystal phase does not crucially depend on the dipolar form of the interaction at long distances \cite{CJB10}. It is interesting that it was shown recently that a superfluid-quasicrystal stripes state with the minimal fivefold rotational symmetry can be realized as the ground state of a Bose-Einstein condensate within a practical experimental scheme \cite{HHSZ18}.

A major theoretical approach to supersolids is via the Gross-Pitaevskii (GP) equation \cite{JPR07}, which is a mean-field approximation for the interparticle interactions and describes a zero-temperature BEC \cite{ESY07}.
Quantum Monte Carlo technique use short range interparticle potential on a lattice to study the supersolid phase of hardcore bosons with nearest-neighbor repulsive interactions on a square supperlattice formed by an external potential \cite{ND17}. The superfluid density has been characterized by extrapolating the imaginary-time diffusion distance of the world lines to infinity in a lattice model and not a continuum system  with out determining the condensate \cite{GB124,GB125}. However, the need for a condensate to explain two-dimensional superfluids is attested in the Kosterlitz-Thouless model, which originally invoked a topological long-range order, rather than on the behavior of a two-point correlation function \cite{KT72}. However, in a subsequent publication, Kosterlitz and Thouless identified the topological order with a nonuniform condensate \cite{KT73}. The need for a nonuniform Bose-Einstein condensate in one- and two-dimensional Bose systems was proved with the aid of the Bogoliubov inequality \cite{MA71,MA19}. In addition, a self-organized, two-dimensional supersolid requires the interaction between the bosons to be nonlocal and of infinitely long-range \cite{MA18}. Therefore, the relationship between superfluids, supersolids and BEC can hold in any dimensional system and so there is no need to replace BEC by a topological order and power-law correlations for finite-temperature superfluids and supersolids in two-dimensional systems.

Extensive Monte Carlo simulations of $^4\textup{He}$ monolayer films adsorbed on weak substrates have been carried out, using various interactions of a helium atom with the substrate, which yield strong evidence that $^4\textup{He}$ will not form a supersolid film on any substrate strong enough to stabilize a crystalline layer \cite{M11}. The specific heat of a two-layer $^4\textup{He}$ film adsorbed on a graphite substrate is estimated as a function of temperature by quantum Monte Carlo simulations where neither the supersolid nor the superfluid hexatic phases are observed \cite{BM20}.

The use of a condensate, whereby a macroscopic number of atoms act in unison, underlies all the theories of superfluids \cite{BP12} and, more recently, all of supersolids \cite{BP12, LMZ17, LLH17}. Studies of the ground state of a dipolar Bose gas with a condensate exhibits droplet crystal states that arrange into a lattice pattern that break rotational symmetry \cite{BB18}. The dipole-dipole interaction is nonlocal and long-ranged \cite {JG14} as required for a supersolid \cite{MA18}.

In this work, we establish that a two-dimensional lattice model cannot by itself, without the aid of nonlocal and of infinitely long-range interparticle potential between the bosons, give rise to a condensate and thus a lattice model per se for supersolids is not attainable.

\section {Symmetry breaking}
Consider the Hamiltonian for an interacting Bose gas
\[
\hat{H}= \int d\textbf{r}\hat{\psi}^{\dag}(\textbf{r}) (\frac{-\hbar^2 }{2m} \nabla^2) \hat{\psi}(\textbf{r}) +  \int d\textbf{r}\hat{\psi}^{\dag}(\textbf{r}) V_{ext}(\textbf{r}) \hat{\psi}(\textbf{r})
\]
\begin{equation}
+ \int d\textbf{r}_{1}\int d\textbf{r}_{2} \int d\textbf{r}_{3} \int d\textbf{r}_{4} \hat{\psi}^{\dag}(\textbf{r}_{1})\hat{\psi}^{\dag}(\textbf{r}_{2})
\end{equation}
\[
\times V(\textbf{r}_{1}, \textbf{r}_{2},\textbf{r}_{3},\textbf{r}_{4}) \hat{\psi}(\textbf{r}_{4})\hat{\psi}(\textbf{r}_{3}),
\]
where $V_{ext}(\textbf{r})$ is an arbitrary, external potential, $V(\textbf{r}_{1}, \textbf{r}_{2},\textbf{r}_{3},\textbf{r}_{4})$ is the two-particle interaction potential, and $\hat{\psi}(\textbf{r})$ and $\hat{\psi}^{\dag}(\textbf{r})$ are bosonic field operators that destroy or create a particle at spatial position $\textbf{r}$, respectively. The two-particle interaction potential $V(\textbf{r}_{1}, \textbf{r}_{2},\textbf{r}_{3},\textbf{r}_{4})$ must satisfy the following general conditions: (i) translational invariance, (ii) Galilean invariance, (iii) identical particles, (iv) rotational invariance, (v) space-reflection invariance, (vi) time-reversal invariance, and (vii) Hermiticity \cite{MA71}. Therefore, in general,

\begin{equation}
V(\textbf{r}_{1}, \textbf{r}_{2},\textbf{r}_{3},\textbf{r}_{4}) = \delta(\textbf{r}_{1} + \textbf{r}_{2} -\textbf{r}_{3} -\textbf{r}_{4}) \langle \textbf{r}_{1} - \textbf{r}_{2}|V| \textbf{r}_{3} - \textbf{r}_{4}\rangle,
\end{equation}
which is referred to as a nonlocal potential. A mathematically simpler potential can be deduced from Eq. (2) if, in addition,

\begin{equation}
\langle \textbf{r}_{1} - \textbf{r}_{2}|V| \textbf{r}_{3} - \textbf{r}_{4}\rangle = \delta(\textbf{r}_{1} - \textbf{r}_{2} -\textbf{r}_{3} +\textbf{r}_{4}) V(|\textbf{r}_{1} - \textbf{r}_{2}|),
\end{equation}
and so
\begin{equation}
V(\textbf{r}_{1}, \textbf{r}_{2},\textbf{r}_{3},\textbf{r}_{4}) = \frac{1}{2} \delta(\textbf{r}_{1}  -\textbf{r}_{3}) \delta(\textbf{r}_{2}  -\textbf{r}_{4}) V(|\textbf{r}_{1} - \textbf{r}_{2}|),
\end{equation}
which is referred to as a local potential.

Macroscopic occupation in the single-particle state  $\psi(\textbf{r})$ result in the non-vanishing \cite{NNB60} of the quasi-average $\psi(\textbf{r}) = <\hat{\psi}(\textbf{r})>$ and so the boson field operator

\begin{equation}
\hat{\psi}(\textbf{r}) =\psi(\textbf{r}) + \hat{\varphi}(\textbf{r}),
\end{equation}
where
\begin{equation}
\hat{\varphi}(\textbf{r}) = \sqrt{\frac{1}{V(D)}}\sum_{\textbf{k}} \hat{ a}_{\textbf{k}} e^{\textbf{k}\cdot\textbf{r}}
\end{equation}
with
\begin{equation}
\psi(\textbf{r}) = \sqrt{\frac{N_{0}}{V(D)}}\sum_{\textbf{k}^\prime} \xi_{\textbf{k}^\prime} e^{i \textbf{k}^\prime \cdot\textbf{r}} \equiv\sqrt{\frac{N_{0}}{V(D)}} f(\textbf{r}) ,
\end{equation}
and
\begin{equation}
\sum_{\textbf{k}^\prime}|\xi_{\textbf{k}^\prime}|^2 =1,
\end{equation}
where $N_{0}$ is the number of atoms in the condensate and $V(D)$ is the D-dimensional ``volume" and $<\hat{\varphi}(\textbf{r})> =0$.  The operator $\hat{\varphi}(\textbf{r})$ has no Fourier components with momenta $\{\textbf{k}^\prime\}$ that are macroscopically occupied and so $\int d\textbf{r} \hat{\varphi}^\dag(\textbf{r}) \psi(\textbf{r}) = 0$. The separation of $\hat{\psi}(\textbf{r})$ into two parts (5) gives rise to the following (gauge invariance) symmetry breaking term in the Hamiltonian (1)

\[
\hat{H}_{symm} =  \int d\textbf{r}_{1} \hat{\varphi}^\dag(\textbf{r}_{1}) \int d\textbf{r}_{2}\int d\textbf{r}_{3}\int d\textbf{r}_{4} \psi^*(\textbf{r}_{2})
\]
\begin{equation}
\times[ V(\textbf{r}_{1}, \textbf{r}_{2},\textbf{r}_{3},\textbf{r}_{4}) + V(\textbf{r}_{2}, \textbf{r}_{1},\textbf{r}_{3},\textbf{r}_{4}) ]\psi(\textbf{r}_{3})\psi(\textbf{r}_{4})+ h. c.
\end{equation}
\[
\equiv \int d\textbf{r}_{1} \hat{\varphi}^\dag(\textbf{r}_{1}) \chi(\textbf{r}_{1}) + h. c.
\]

There are sixteen terms resulting from the substitution of (5) into (1). These terms can be classified according to the number of factors of $\psi(\textbf{r})$ and/or $\psi^*(\textbf{r})$. The symmetry breaking term (9) follows from the terms with three factors of $\psi(\textbf{r})$ and/or $\psi^{*}(\textbf{r})$. All other terms, except those with four factors of $\psi(\textbf{r})$ and/or $\psi^{*}(\textbf{r})$ and four factors of $\hat{\varphi}(\textbf{r})$ and/or $\hat{\varphi}^{\dag}(\textbf{r})$, represent interactions between the particles in the condensate and those not in the condensate, that is, particles going in and out of the condensate via these interactions.

The presence of this nonzero $\hat{H}_{symm}$ in the Hamiltonian gives rise to further macroscopic occupation in states other than the original state given by $\psi(\textbf{r})$ and so the condensate wave function $\psi(\textbf{r})$ gets modified by augmenting the single-particles states where macroscopic occupation occurs. In such a case, macroscopic occupation in the state $b$ would give rise to macroscopic occupation in the states $a$, such that $a\neq b$, whenever  the matrix element $<ab|\hat{V}|bb>$  of the potential $\hat{V}$, which is the last term in Eq. (1), does not vanish.  For instance, macroscopic occupation only in the single-particle state with momentum $\textbf{p}$, which corresponds to a uniform condensate, does not give rise to macroscopic occupation in any other momentum state since the matrix element in the momentum representation $<\textbf{q}\textbf{p}| \hat{V}|\textbf{p}\textbf{p}>$ vanishes by momentum conservation unless $\textbf{q}= \textbf{p}$. This consistency proviso requires that the correct condensate wave function $\psi(\textbf{r})$ corresponds to that which gives rise to no symmetry breaking term in the Hamiltonian. That is to say, $\hat{H}_{symm}$ vanishes for the correct condensate wave function $\psi(\textbf{r})$.

For instance, macroscopic occupation in the single-particle states with momenta $\textbf{k}, \textbf{k} \pm\textbf{q}_{1}, \textbf{k} \pm\textbf{q}_{2}$, where $\textbf{q}_{1}\times \textbf{q}_{2}\neq \textbf{0}$, gives rise, with the aid of the symmetry breaking term $\hat{H}_{symm}$ and owing to linear momentum conservation, to additional macroscopic occupation in single-particle momenta states. Therefore, for $\hat{\varphi}^\dag(\textbf{r})$ to be orthogonal to both $\psi(\textbf{r})$ and $\chi(\textbf{r})$ and so $\hat{H}_{symm}=0$, one must have macroscopic occupation in all the momentum states $\textbf{k} + n_{1}\textbf{q}_{1} + n_{2}\textbf{q}_{2}$, with $n_{1}, n_{2} = 0, \pm 1, \pm 2, \cdots.$   Accordingly,
\[
\psi_{\textbf{k}}(\textbf{r}) = \sqrt{\frac{N_{0}}{V(D)}} \sum_{n_{1},  n_{2} = - \infty}^{\infty}  \xi_{\textbf{k}+ n_{1} \textbf{q}_{1} + n_{2} \textbf{q}_{2} } \hspace{0.1in} e^{i (\textbf{k}+ n_{1} \textbf{q}_{1} + n_{2} \textbf{q}_{2} )\cdot\textbf{r}}
\]
\begin{equation}
\equiv  e^{i \textbf{k}\cdot \textbf{r}} u_{\textbf{k}}(\textbf{r}).
\end{equation}

Note that the BEC (10) generates a real-space, crystalline distribution of atoms since $u_{\textbf{k}}(\textbf{r}) = u_{\textbf{k}}(\textbf{r} +\textbf{t}_{m})$ for any primitive lattice translation vector $\textbf{t}_{m}$ with $e^{\textbf{q}_{i}\cdot \textbf{t}_{m}} =1$ for $i=1, 2$ and so (10) is of the Bloch form. For 2D, the vector $\textbf{t}_{m} = m_{1} \textbf{a} + m_{2} \textbf{b}$, where $m_{1}$ and $m_{2}$ can take all integer values and $\textbf{a}$ and $\textbf{b}$ are the edges of the unit cell, which form parallelograms given by the five Bravais lattices. Note that $\textbf{a} \cdot \textbf{q}_{1} =2 \pi$,  $\textbf{a} \cdot \textbf{q}_{2}= 0$,  $\textbf{b} \cdot \textbf{q}_{1} = 0$, and  $\textbf{b} \cdot \textbf{q}_{2}=2 \pi$. The previous case refers to a two-dimensional lattice, the case for a three-dimensional lattice would require a third vector $\textbf{q}_{3}$ such that $\textbf{q}_{1}\times (\textbf{q}_{2} \times \textbf{q}_{3})\neq \textbf{0}$ with $\textbf{t}_{m} = m_{1} \textbf{a} + m_{2} \textbf{b}+m_{3} \textbf{c}$.

It is important to remark that the above Bose-Einstein condensate (10) is appropriate for two-dimensional supersolids. Note that the existence of a Bose-Einstein condensate in superfluids for $D\leq 2$ requires points of accumulations of condensates in dense sets of single particles momentum states \cite{MA71}. This is quite analogous to generating a Fourier integral from a Fourier series.

\section{Lattice Models}

The Bloch form of the condensate (10) imposes no conditions on the interparticle potential for 2D; however, the formation of a supersolid in $D\leq 2$ and finite temperature $T>0$ requires the interparticle potential to be infinitely long-range and nonlocal \cite{MA19}. The replacement (5) in the Hamiltonian (1) was considered only for the interparticle potential in (1) in order to generate a consistent condensate via the the symmetric Hamiltonian $\hat{H}_{symm}$ given in (9) \cite{MA19}.

We now consider the replacement (5) in the term in (1) associated with the external, local potential $V_{ext}(\textbf{r})$. One obtains the symmetry breaking Hamiltonian
\begin{equation}
\hat{H}_{symm}^{(ext)}= \int\textup{d\textbf{r}} \hat{\varphi}^\dag(\textbf{r})V_{ext}(\textbf{r})\psi(\textbf{r}) + h.c.
\end{equation}

Consider the local, finite two-dimensional lattice potential,
\begin{equation}
V_{ext}(\textbf{r})=\frac{1}{(2\pi)^2}\sum_{\textbf{k}} g(\textbf{k}) \sum_{m_{1},m_{2}=-M}^{M} e^{-i\textbf{k}\cdot(\textbf{r}-m_{1}\textbf{a}-m_{2}\textbf{b})}
\end{equation}
where $g(\textbf{k})$ is the Fourier transform, \textbf{a} and \textbf{b} are arbitrary two-dimensional vectors in the x-y plane, and $\textbf{a}\times \textbf{b}\neq \textbf{0}$. One obtains that
\begin{equation}
\hat{H}_{symm}^{(ext)}=\int \textup{d\textbf{r}} \hat{\varphi}^\dag(\textbf{r}) V_{ext}(\textbf{r})\psi(\textbf{r})+h.c.
\end{equation}
\[
=\frac {\sqrt{N_{0}}}{V(D)} \sum_{\textbf{k}_{1}, \textbf{k}_{2}\neq \textbf{k}_{1} }\hat{a}_{\textbf{k}_{1}} ^{\dag} \xi_{\textbf{k}_{2}}g(\textbf{k})\frac{ \sin[(\textbf{k}\cdot \textbf{a})(M+1/2)]}{\sin[(\textbf{k}\cdot \textbf{a})/2]}
\]
\[
\times\frac{ \sin[(\textbf{k}\cdot \textbf{b})(M+1/2)]}{\sin[(\textbf{k}\cdot \textbf{b})/2]} +h.c.,
\]
where $ \textbf{k} \equiv\textbf{k}_{2} -\textbf{k}_{1}$, which follows with the aid of (5), (6), (12), and
\begin{equation}
\sum_{m=-M}^{M} e^{imx} = \frac{\sin[x(M+1/2)]}{\sin[x/2]} \rightarrow  2\pi\delta(x) \hspace{.2in} ( M\rightarrow \infty).
\end{equation}
Note that $\textbf{k}_{1} \neq \textbf{k}_{2}$, that is, $\textbf{k}\neq \textbf{0}$,  since $\textbf{k}_{2}$ is in the condensate and $\textbf{k}_{1}$ is not. Therefore, $\hat{H}_{symm}^{(ext)}$ vanishes in the macroscopically large lattice limit and so one cannot generate a two-dimensional supersolid at finite temperature from an external lattice potential. A two-dimensional supersolid at finite temperature can be generated via long-range, nonlocal potentials provided by the interparticle interaction which results in self-organization, much as Wigner crystallization or Wigner lattice, electrons moving in a uniform background of positive charge that restore electric neutrality \cite{EW34}.

For instance, a recent article purports to show the existence of a quasi-two-dimensional supersolid at zero temperature with the external interaction of a He atom and a graphite surface \cite{GB124}. The authors consider a trial wave function which does not give rise to a Bose-Einstein condensate but, in fact, the trial wave function actually vanishes for a two-dimensional, infinite lattice.

The corrected trial wave function used by the authors \cite{GB125}, is
\begin{equation}
\Phi_{2}(\textbf{r}_{N+1},\cdots\textbf{r}_{N}) = \prod_{i=N_{1}+1}^{N} \Psi_{2}(z_{i})
\end{equation}
\[
\times \prod_{I=1}^{N_{2}}\Bigg{[}\sum_{i=N_{1}+1}^{N} \exp{\{-a_{2}|\textbf{r}_{i}-\textbf{r}_{I}^{(2)}|^2\}}\Bigg{]},
\]
where $N_{1}$ if the number of atoms in layer 1, $N_{2}$ is both the number of atoms in layer 2 and the number of lattice points of the solids, and $N_{1}+N_{2}=N$. Therefore, no vacancies were considered in any solid. The vectors $\textbf{r}_{i}$ and $\textbf{r}_{I}^{(2)}$ are both in the $x$-$y$ plane. The latter two-dimensional plane is where the supersolid would occur.

The important term is the second product in the right-hand-side (RHS) of (15). The sum of terms in (15) are of the form
\[
\sum_{i,j,\cdots k}\exp{\{-a_{2}|\textbf{r}_{N_{1}+i}-\textbf{r}_{N_{1}+1}^{(2)}|^2\}} \exp{\{-a_{2}|\textbf{r}_{N_{1}+j}-\textbf{r}_{N_{1}+2}^{(2)}|^2\}}
\]
\begin{equation}\cdots \exp{\{-a_{2}|\textbf{r}_{N_{1}+k}-\textbf{r}_{N_{1}+N{2}}^{(2)}|^2\}}
\end{equation}
consisting of $N_{2}$ terms, where $i,j, \cdots k$ take values $1,2,3, \cdots$ that can be repeated.

It is clear that in order to consider an infinite two-dimensional lattice we will have to start with a finite lattice and then consider the limit as the size of the lattice goes to infinity. We consider a strictly two-dimensional layer generated by the primitive lattice translation vector $\textbf{t}_{m}= m_{1}\textbf{a}+m_{2}\textbf{b}$ with integers $\{m_{i}\} = 0, \pm 1, \pm 2 \cdots$, where $\textbf{a}$ and $\textbf{b}$ are arbitrary two-dimensional vectors in the  x-y plane. The following factor is present in every term in (16), viz.,
\begin{equation}
\exp{\{-a_{2}|\textbf{r}_{N_{1}+1}^{(2)}|^2\}} \exp{\{-a_{2}|\textbf{r}_{N_{1}+2}^{(2)}|^2\}}\cdots \exp{\{-a_{2}|\textbf{r}_{N_{1}+N_{2}}^{(2)}|^2\} }.
\end{equation}
The factor in the exponent in (17) is
\begin{equation}
|\textbf{r}_{N_{1}+1}^{(2)}|^2 +|\textbf{r}_{N_{1}+2}^{(2)}|^2\cdots |\textbf{r}_{N_{1}+N_{2}}^{(2)}|^2=\sum_{m_{1},m_{2}=-M}^{M}|m_{1}\textbf{a}+m_{2}\textbf{b}|^2
\end{equation}
\[
 =\frac{1}{3}(a^2+b^2)M(M+1)(2M+1)^2,
\]
since $\sum_{m=-M}^{M}m=0$ and $\sum_{m=-M}^{M}m^2=\frac{1}{3}M(M+1)(2M+1)$. The sum in (16) contains $N_{2}^2 = e^{4 \ln(2M+1)}$ terms. Therefore, in the limit of an arbitrary, infinite two-dimensional lattice, the trial wave function vanishes. Obviously, then there is no Bose-Einstein condensate either with no off-diagonal long-range order \cite{Y62}.

\section{Summary and discussion}

We have established that a two-dimensional supersolid at finite temperatures arises from infinitely, long-range potentials and not via an external lattice potential. However, a three-dimensional supersolid places no restriction on the interparticle potential. We present a correct form for the trial wave function of a two-dimensional supersolid at any temperatures that satisfies the Bloch form.

\bibliography{basename of .bib file}

\end{document}